\documentclass[aps,prl,twocolumn,showpacs,groupedaddress]{revtex4}
\usepackage{graphicx}
\begin{document}
\title{Quasiparticle Evolution and Pseudogap Formation in V$_2$O$_3$: An Infrared Spectroscopy Study}

\date{\today}
\author{L. Baldassarre$^{1}$, A. Perucchi$^{1,2}$, D. Nicoletti$^{1}$, A. Toschi$^{3}$, G. Sangiovanni$^{3}$, K. Held$^{3}$, M.~Capone$^{4}$, M. Ortolani$^{5}$, L. Malavasi$^{6}$, M. Marsi$^{7}$, P. Metcalf$^{8}$ P. Postorino$^{1}$, and S. Lupi$^{1}$}
\affiliation{$^{1}$CNR-INFM COHERENTIA and Dipartimento di Fisica,  Universit\`{a} di Roma  ``La Sapienza'',  Piazzale Aldo Moro 2, I-00185 Roma, Italy}
\affiliation{$^2$ Sincrotrone Trieste S.C.p.A., S.S. 14 Km 163.5, in Area Science Park, 34012 Basovizza Trieste, Italy}
\affiliation{$^3$Max-Planck Institut f\"ur Festk\"orperforschung, Heisenbergstr. 1, D-70569 Stuttgart, Germany}
\affiliation{$^4$SMC, CNR-INFM and Dipartimento di Fisica,  Universit\`{a} di Roma ``La Sapienza'', Piazzale Aldo Moro 2, I-00185 Roma, Italy and ISC-CNR, Via dei Taurini 19, Roma, Italy}
\affiliation{$^5$ Istituto di Fotonica e Nanotecnologie, IFN-CNR, Via Cineto Romano 42, 00156 Roma, Italy}
\affiliation{$^6$ Dipartimento di Chimica Fisica ``M. Rolla'', INSTM and IENI-CNR, Universit\`a di Pavia, Viale Taramelli 16, I-27100 Pavia, Italy}
\affiliation{$^7$ Laboratoire de Physique des Solides UMR 8502- Universit\'e Paris-Sud, B\^at. 510 - 91405 Orsay cedex}
\affiliation{$^8$ Department of Chemistry, Purdue University, West Lafayette, IN 47907, USA.}

\begin{abstract}
The infrared conductivity of V$_2$O$_3$ is measured in the whole phase diagram.
Quasiparticles appear above the N\'eel temperature $T_N$ and eventually disappear further enhancing the 
temperature, leading to a pseudogap in the optical spectrum above 425 K. 
Our calculations demonstrate that this loss of coherence can be explained only if the temperature dependence 
of lattice parameters is considered. 
V$_2$O$_3$ is therefore effectively driven from the ``metallic'' to the ``insulating'' side of the Mott 
transition as the temperature is increased.
\end{abstract}

\maketitle

Conventional metals are characterized by well-defined long-lived quasiparticles (QP) that exist up to a coherence temperature $T_{coh}$ of the order of the Fermi temperature. 
Due to the large value of $T_{coh}$ in comparison with experimentally accessible temperatures, one cannot observe the loss of coherence and of spectral weight expected when $T \gtrsim T_{coh}$. 
On the other hand, in strongly
correlated electron systems the coherence scale is generally reduced. 
This is particularly true in the proximity to the Mott transition, where the disappearance of the QP across $T_{coh}$ \cite{georges_rmp} is therefore expected to be observable. 
However, a direct experimental observation of this phenomenon is still lacking in what is considered the 
prototype of Mott-Hubbard systems: vanadium sesquioxide (V$_2$O$_3$).
In particular, there is no characterization, either with optics or photoemission\cite{notaARPES},
of what happens when QP eventually lose their coherence as temperature is increased.

\begin{figure}[h]
\includegraphics[width=8.5cm]{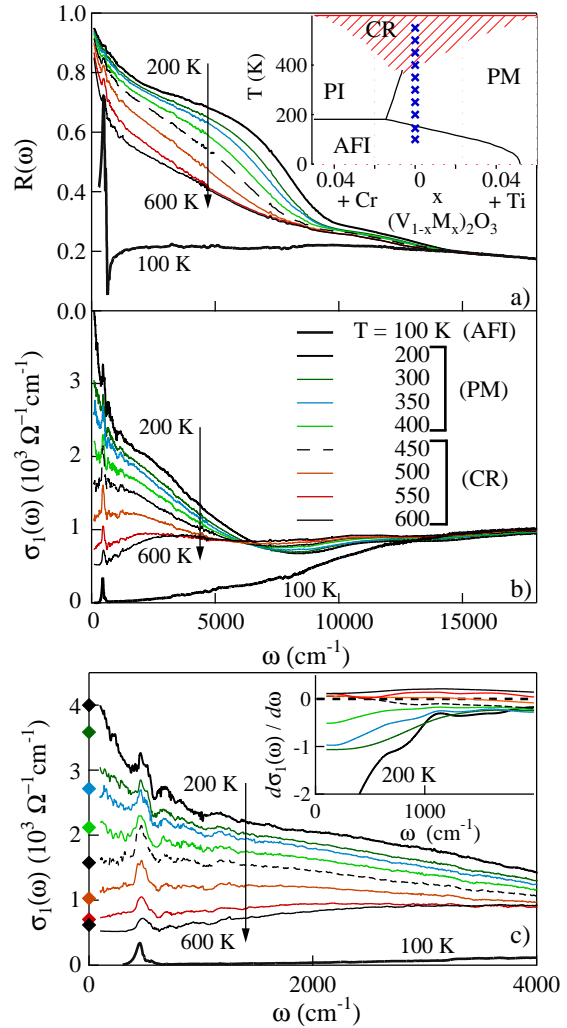}
 \caption{(Color online) a) Near-normal incidence reflectivity of V$_2$O$_3$ shown at selected $T$ over the 0 $\div$ 18000 cm$^{-1}$ range. The phase diagram upon doping with Cr and Ti \cite{mcwhan73} is recalled in the inset. Markers indicate the measurements performed in the present paper. b) Optical conductivity obtained from the reflectivity in a), plotted in the same spectral range for the same $T$. c) Optical conductivity is shown between 0 and 4000 cm$^{-1}$ to stress the low-energy behavior. Diamonds represent the DC conductivity values as measured by four probe resistivity technique. In the inset of c) $d\sigma_1/d\omega$ is plotted as a function of frequency up to 2000 cm$^{-1}$.} \label{fig1}
\end{figure}

V$_2$O$_3$ presents a rich phase diagram shown in the inset of Fig.\ref{fig1}a \cite{mcwhan73}.
 On increasing $T$ the system undergoes a transition from a low-$T$ insulating antiferromagnetic phase (AFI), with a monoclinic crystal structure, to a high-$T$ paramagnetic correlated metallic phase (PM) with a corundum tetragonal structure. The insulating phase shows a high resistivity \cite{Kuwamoto,mcwhan73}, indicating the absence of propagating QP excitations.  The resistivity is reduced by several orders of magnitude when crossing the insulator to metal transition (MIT) at $T_N$ = 160 K \cite{Morin}.
Above 400K a badly conducting crossover regime (CR) sets in, which has never been extensively investigated.

In this work we present measurements extending over a temperature range much larger than in 
previous spectroscopic investigations (performed at 300, 170 and 70 K  \cite{thomas,remeika}).
This analysis is made possible by studying the infrared reflectivity R($\omega$) of a high quality
single crystal of V$_2$O$_3$ at near-normal incidence. 
Our work represents the first systematic optical study of all the regions (AFI, PM and CR) of the
phase diagram, and it allows us to follow the development of the QP emerging at $T_N$ from the AFI
insulating phase, their evolution into the metallic state for increasing temperature and their 
progressive disappearance above $T_{coh} \approx 425$ K, as the system enters the CR.

The extension to high temperature leads us to the first experimental evidence of the opening of 
a pseudogap in this material. This is characterized by a strong loss of spectral weight ($SW$) at low 
frequency and a downturn in the optical conductivity for $\omega \rightarrow 0$ (note that an analogous 
downturn has been reported below 300 K for strongly correlated systems such as SrRuO$_3$\cite{kostic} and, 
at higher temperatures, for La$_{2-x}$Sr$_x$CuO$_4$\cite{downturn_cuprates}).

We also perform calculations with LDA+DMFT (local density approximation combined with dynamical mean-field 
theory\cite{metdmft,georges_rmp}) below and above $T_{coh}$. 
We find that the key effect behind the pseudogap formation is the temperature dependence of 
lattice parameters \cite{mcwhan69}, never considered in previous calculations. 
The change is small if measured on an absolute scale, but it has a surprising outcome because of the 
proximity to the Mott transition.
At high temperatures, V$_2$O$_3$ lies in facts on the ``insulating'' side of the phase diagram, as
if, above $T_{coh}$, one had to bend to the left the blue-crosses straight line in the inset 
of Fig.\ref{fig1}a.

The R($\omega$) data  were collected on a high quality single crystal\cite{harrison} between 100 and 21000 cm$^{-1}$ with a Michelson interferometer and for temperatures between 100 and 600 K, completely covering the AFI, PM and CR phases. Reference was taken on a metallic film (gold or silver, depending on the spectral range) evaporated $in$ $situ$ over the sample surface. The real part of the optical conductivity, $\sigma_1(\omega)$, has been calculated through Kramers-Kronig transformations. Standard extrapolation procedures \cite{dressel} were adopted for the insulating or metallic (crossover) phases at low frequency (constant or Hagen-Rubens extrapolation), while a high frequency tail was merged to our data, as in Ref. \onlinecite{thomas}. 

The reflectivity at selected $T$ in the AFI, PM and the CR phases is shown in Fig. \ref{fig1}a.
At $T=100$ K (i.e., well below $T_N$) and at 150 K- not shown -  R($\omega$) exhibits an insulating behavior with a phonon peak (not discussed here) around 500 cm$^{-1}$. R($\omega$) is nearly flat up to about 10000 cm$^{-1}$ where a weak electronic absorption band appears \cite{rif_AFI}. For $T>$ 200 K instead, the reflectivity is metallic-like (i.e., R$\rightarrow 1$ for $\omega\rightarrow 0$) and decreases over the whole infrared range as temperature increases. Almost no temperature dependence is found above 15000 cm$^{-1}$ where all the reflectivity curves eventually merge.

A close inspection of Fig.~\ref{fig1}b reveals in the three regimes different behaviors for $\sigma_1(\omega)$. In the AFI state, $\sigma_1(\omega)$ -in good agreement with previous data \cite{thomas,remeika}- shows a well defined charge gap in the infrared. The absorption in the gap rises with increasing frequency up to a broad feature in the visible.
This is the Mott gap between the Hubbard bands of the strongly correlated AFI \cite{georges_rmp,rozenberg95}.
On crossing $T_N$ an abrupt filling of the gap is induced and $\sigma_1(\omega)$ shows, below 1 eV, a metallic absorption due to the appearance of QP, in good agreement with what observed for resistivity \cite{Kuwamoto,Morin} and specific heat \cite{mcwhan71}. As $T$ is raised, $\sigma_1(\omega)$ presents a strong temperature dependence (see Fig.~\ref{fig1}b), with a huge transfer of $SW$ from low to high frequency through an isosbestic point at about 6000 cm$^{-1}$.
In addition to the metallic term, a broad band (MIR) is observed in the mid-infrared. 
This feature was assigned \cite{thomas}  to the optical transitions from Hubbard bands to the states around the Fermi 
energy\cite{rozenberg95,georges_rmp}. 

We have also performed four probe resistivity measurements in the 200-600K range \cite{Malavasi}, in order to follow the development of the metallic absorption and its evolution with $T$. The behavior of $\sigma_1(\omega)$ at low frequency has been compared with DC conductivity (diamonds in Fig.~\ref{fig1}c). The $\omega$ $\rightarrow$ $0$ limit of $\sigma_1(\omega)$ and the DC conductivity are in excellent agreement at almost all temperatures.
This agreement and the overall decreasing behavior of $\sigma_1(\omega)$ ($d\sigma_1/d\omega<0$), clearly observable for $T < 400$ K (see inset of Fig. \ref{fig1}c), suggests the presence of a low-frequency QP contribution between 200 and 400 K.
For $T=450$ K, the low-frequency $\sigma_1(\omega)$ starts to show a downturn, $d\sigma_1/d\omega>0$ in a frequency range that increases with increasing $T$ (inset of Fig.~\ref{fig1}c). We associate this change of sign with the disappearance of QP. 
On increasing $T$ from 450 to 600 K, the downturn eventually transforms into a pseudogap at low frequency, while the MIR 
broadens.
The pseudogap formation in $\sigma_1(\omega)$ is in agreement with an anomalous enhancement of the DC resistivity of 
V$_2$O$_3$ when entering the CR \cite{Malavasi,Kuwamoto,mcwhan73}. Our optical measurements and, in particular, the changes 
of $\sigma_1(\omega)$ observed at low frequencies between $400$ and $450$ K (see inset of Fig. \ref{fig1}c) lead to an optical 
estimate of $T_{coh} \approx 425$K in V$_2$O$_3$.

To quantify the QP term just above $T_N$, and its reduction on increasing $T$, we have calculated the spectral weight $SW$ using the restricted $f$ sum-rule \cite{dressel}
$SW(\Omega, T)= \frac{2mV_{V}}{\pi e^2}\int^{\Omega}_0 \sigma_1(\omega,T)d\omega$ where m is the mass of carriers, $e$ is the electron charge and $V_{V}$ is the volume per V ion.

\begin{figure}[h]
\includegraphics[width=8 cm]{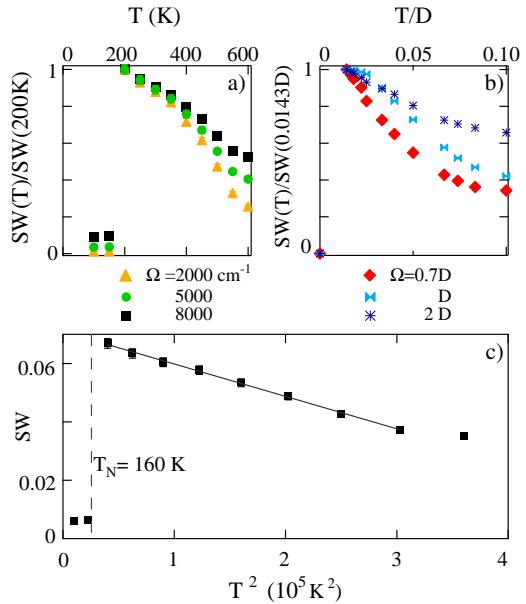}
 \caption{(Color online) Normalized $SW$ for three different cutoffs $\Omega$ as a function of $T$ in: a) experiments and in b) a single-band Hubbard model. The normalization of the theoretical data coincides with the experimental one if $D=1$ eV.  c) $SW$ integrated up 8000 cm$^{-1}$ as a function of $T^2$, showing a linear behavior.} \label{fig2}
\end{figure}

The temperature behavior of $SW$ is reported for three cutoff energies $\Omega$  in Fig.~\ref{fig2}a, normalized at the 200 K value. $SW$ is nearly constant in the AFI state and shows at $T_N$ a large, discontinuous jump,  due to the appearance of QP . However, on further increasing $T$, $SW(T)$ decreases. At the lower cutoff ($\Omega$= 2000 cm$^{-1}$) the dramatic loss of $SW$ indicates the loss of coherence mainly of the metallic term while at  the highest $\Omega$ value (8000 cm$^{-1}$), the decrease of $SW$ is roughly due to both the QP contribution and the MIR band.   Therefore the loss of $SW$ mainly reflects the lowering of the QP contribution, and, eventually, its disappearance.

As shown in Fig.~\ref{fig2}c, for 200 K $\leq T \leq$ 550 K the $SW$ is proportional to $T^2$, a characteristic behavior of Fermi liquids.
We repeat the analysis introduced for the cuprates \cite{Ortolani, Toschi}, describing the $SW$ as $SW(\Omega, T) = SW(\Omega,0)-B(\Omega)T^2$, where $B(\Omega)$ depends on the effective QP bandwidth, providing an estimate of the strength
of correlation.
To single out the free-particle and MIR contributions to the $SW$ we choose as a cutoff $\Omega$=  8000 cm$^{-1}$, where $\sigma_1(\omega)$ has a minimum. The value of $B(\Omega)$ does not however significantly change if $\Omega$ is chosen up to 15000 cm$^{-1}$, as far as the discussion in the present paper is concerned. The same value of $B(\Omega)$, within the experimental error, is obtained also by subtracting the AFI $\sigma_1(\omega)$ from those in the PM state before integrating. 

Defining $b(\Omega)=B(\Omega)/SW(\Omega,0)$ we obtain, for $\Omega$=  8000 cm$^{-1}$, $b=1.6\times 10^{-6}$ K$^{-2}$. 
This value indicates that stronger correlation takes place in V$_2$O$_3$ with respect, for example, to LSCO cuprate ($b= 2.5 \times 10^{-7}$  K$^{-2}$) \cite{Ortolani}. 
$b(\Omega)$ is instead comparable with that of other correlated compounds as Nd$_{0.905}$TiO$_3$  ($b= 2.2 \times 10^{-6}$  K$^{-2}$) \cite{timusk}. All those values are, however, much higher than that in a conventional metal such as gold ($b = 1.3 \times 10^{-8}$  K$^{-2}$) \cite{Ortolani}.

In the following we try to establish more firmly the role of correlations in determining the observed phenomena.
Even if V$_2$O$_3$ is usually considered the prototype of Mott-Hubbard systems, a number of indications suggest that its properties cannot be consistently understood in terms of a single-band Hubbard model, and multiband effects as well as realistic physical aspects of the system need to be considered\cite{park,karsten,oka}.
Using the same effective $U$ values as in \cite{rozenberg95} ($U=2D$ for the PM and $U=4D$  for the AFI phase, where $D$ is the half bandwidth), even the single band half-filled Hubbard model provides a correct estimate of the decrease of $SW$ in the considered temperature interval, ascribing it to correlation effects (see Fig. \ref{fig2}b). This reduction ranges from 40 to 70\% for the different cut-offs \cite{notadop}.
However, the $T$ scale is off by a factor of two if $D=0.4$ eV is chosen to reproduce the correct size of the Mott gap in the single band Hubbard model\cite{rozenberg95, merino}.
We thus confirm that a single-band model is able to mimic the physics of V$_2$O$_3$ but only using an 
``ad hoc'' prescription. 

\begin{figure}[ht]
\includegraphics[width=9cm]{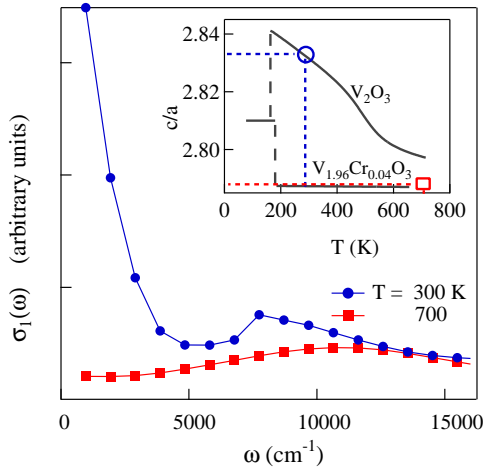}
\caption{(Color online) LDA+DMFT optical conductivity for V$_2$O$_3$ at two different temperatures. In the inset the temperature dependence of the ratio $c/a$ is reported (Data from Ref. \onlinecite{mcwhan69}). For V$_2$O$_3$  the change of  $c/a$ observed as a function of $T$ reflects in a dramatic change of the low-frequency optical spectral, i.e. the disappearing of the metallic peak.} \label{LDA_cond}
\end{figure}

We present here a more realistic treatment of strong correlations in V$_2$O$_3$. 
Extending a previous LDA+DMFT calculation\cite{karsten},  we compute the first 
theoretical optical conductivity for V$_2$O$_3$\cite{nota_prob}. In this approach the realistic band 
structure obtained by LDA is used to build up a Hamiltonian which is then solved by DMFT, using quantum 
Monte Carlo (QMC) as an impurity solver.

In Fig. \ref{LDA_cond} we show $\sigma_1(\omega)$ computed with LDA+DMFT at $T$=300 and 700K, 
respectively below and above $T_{coh}$.
Remarkably, our data are in nice agreement with the experimental result of Fig.\ref{fig1}.
In particular we reproduce the observed high-temperature pseudogap.
Had we used the standard input (i.e., the lattice parameters at 300 K) for LDA+DMFT, the optical 
conductivity at 700 K would have displayed a Drude-like peak rather than a well formed pseudogap
as in Fig.\ref{LDA_cond} (red curve with solid squares).
The crucial ingredient of our calculation is the change in lattice parameters of V$_2$O$_3$ 
with increasing temperature, something always overlooked in any previous theoretical 
analysis.
Early studies\cite{mcwhan69} show indeed that the lattice parameters and the atomic 
positions of undoped V$_2$O$_3$ markedly change around 450 K, as opposed to the Cr-doped 
compound whose lattice parameters are nearly constant in temperature.
In particular, above 450K the structure of V$_2$O$_3$ rapidly approaches that of 
insulating V$_{1.96}$Cr$_{0.04}$O$_3$ (i.e. its LDA bandwidth slightly shrinks) as shown by the ratio $c/a$ between the $c$- and 
$a$-axis lattice constant from Ref. {\onlinecite{mcwhan69}} (inset of Fig.\ref{LDA_cond}).
Such effect is taken into account in our calculations in the following way: 
We fix $U$=5 eV\cite{karsten, panaccione} and use, at $T$=300 K, the corresponding V$_2$O$_3$ crystal structure (blue circle in the inset of Fig. \ref{LDA_cond}); for $T$=700 K we use instead  V$_{1.962}$Cr$_{0.038}$O$_3$ as an input, which has almost the same lattice constants and a very similar $c/a$ ratio  as that of V$_2$O$_3$ at the same $T$ (see red box in the inset of Fig. \ref{LDA_cond}).

%In Fig. \ref{LDA_cond} we report that, in nice agreement with the experimental data (see Fig. \ref{fig1}b), $\sigma_1(\omega)$ displays at $T=300$ K a well-defined metallic peak while at $T=700$ K the metallic peak is completely gone.
%At low frequencies $\sigma_1(\omega)$ has a positive slope and a non-vanishing optical weight because of the finite temperature.

In conclusion, we measured the optical properties of V$_2$O$_3$ in the whole phase 
diagram. 
We observed a discontinuous onset of the QP contribution above the MIT and the opening
of a pseudogap above $T_{coh} \approx 425$ K. 
This result also calls for a confirmation from high-temperature photoemission experiments, 
which are still lacking for V$_2$O$_3$.
LDA+DMFT calculation nicely reproduces our experimental data. However, the opening of a 
pseudogap above $T_{coh}$ can only be obtained including the change of the 
crystallographic parameters with temperature.
Atomic or structural disorder does not seem to play a fundamental role here, in contrast 
to what is expected for doped compounds\cite{mlaad}.   
We therefore attribute the pseudogap formation in V$_2$O$_3$ to the temperature dependence 
of lattice constants, which gives a small shrinking of the LDA bandwidth upon heating.
The effect is however dramatic, as the change in the lattice parameters effectively drives 
V$_2$O$_3$ into the ``insulating'' side of the Mott transition.

\acknowledgments{We are indebted to O.~K.~Andersen for pointing out the crucial role of temperature dependence of the crystallographic parameters. We also thank P.~Calvani, C.~Castellani, O.~Gunnarsson, E.~Koch and E.~Pavarini for comments and useful discussions}


\begin{thebibliography}{99}
\bibitem{georges_rmp} A. Georges, 
G. Kotliar, W. Krauth and M.J. Rozenberg,  {\itshape Rev. Mod. Phys.} \textbf{68},13 (1996).

\bibitem{notaARPES} Hitherto the CR phase has been studied by photoemission only coming from the PI insulating phase by S.-K. Mo, H.-D. Kim, J. W. Allen, G.-H. Gweon, J. D. Denlinger, J.-H. Park, A. Sekiyama, A. Yamasaki, S. Suga, P. Metcalf, and K. Held,  {\itshape Phys. Rev. Lett.} \textbf{93}, 076404 (2004).


\bibitem{mcwhan73}  D. B. McWhan, A. Menth, J. P. Remeika, W. F. Brinkman, and T. 
M. Rice, {\itshape Phys. Rev. B} \textbf{7}, 1920 (1973).


\bibitem{Kuwamoto} H. Kuwamoto, J. M. Honig, and J. Appel   Phys. Rev. B \textbf{22}, 2626 (1980),  and P. Limelette, A. Georges, D. Jerome, P. Wzietek, P. Metcalf, and J. M. Honig, Science  {\bf{302}}, 89  (2003).

\bibitem{Morin} F. Morin, Phys. Rev. Lett. \textbf{3}, 34 (1959).

\bibitem{thomas} G.A. Thomas, D.H.Rapkine, S.A. Carter, A.J. Millis, T.F. Rosenbaum, P. Metcalf, and J.M. Honig, {\itshape Phys. Rev. Lett.} \textbf{73}, 1529 (1994). 

\bibitem{remeika}  A.S. Barker and J.P. Remeika,  Solid State Commun. \textbf{8}, 1521 (1970).

\bibitem{kostic} P. Kostic,  Y. Okada, N. C. Collins, Z. Schlesinger, J. W. Reiner, L. Klein, A. Kapitulnik, T. H. Geballe, and M. R. Beasley, Phys. Rev. Lett. {\bf 81}, 2498 (1998).

\bibitem{downturn_cuprates} K. Takenaka, J. Nohara, R. Shiozaki, and S. Sugai, Phys. Rev. B \textbf{68}, 134501 (2003).

\bibitem{metdmft} W. Metzner and D. Vollhardt, Phys. Rev. Lett. {\bf 62}, 324, (1989).

\bibitem{harrison} Growing conditions are reported by H. R. Harrison {\it et al.}, Mater. Res. Bull. \textbf{15}, 571 (1980).

\bibitem{dressel}
M. Dressel and G. Gr\"uner, in Electrodynamics of Solids, Cambridge University Press (2002).

\bibitem{rif_AFI} While $\sigma_1(\omega)$ in the AFI phase has been previously published, to our knowledge this is the first time that R($\omega$) in the same phase  is reported.

\bibitem{mcwhan71} D. B. McWhan,  J. P. Remeika, T. M. Rice, W. F. Brinkman, J. P. Maita, and A. Menth , Phys. Rev. Lett. \textbf{27}, 941 (1971). 

\bibitem{rozenberg95} M.J.Rozenberg, G.Kotliar, H.Kajueter, G.A.Thomas, D.H.Rapkine, 
J.M.Honig, P.Metcalf, {\itshape Phys. Rev. Lett.} \textbf{75}, 105 (1995). 
\bibitem{Malavasi} L. Malavasi, L. Baldassarre {\it et al.},  {\itshape unpublished} (2007).

\bibitem{Ortolani} M. Ortolani, P. Calvani and S. Lupi, Phys. Rev. Lett., \textbf{94}, 067002 (2005).

\bibitem{Toschi} A. Toschi,  M. Capone, M. Ortolani, P. Calvani, S. Lupi, and C. Castellani Phys. Rev. Lett., {\bf 95}, 097002 (2005).

\bibitem{timusk} J. Yang, J. Hwang, T. Timusk, A. S. Sefat, and J. E. Greedan, Phys. Rev. B, {\bf 73}, 195125 (2006).

\bibitem{park} J.-H. Park, L. H. Tjeng, A. Tanaka, J. W. Allen, C. T. Chen, P. Metcalf, J. M. Honig F. M. F. de Groot and G. A. Sawatzky, Phys. Rev. B, {\bf 61}, 11506 (2000).


\bibitem{karsten} K. Held, G. Keller, V. Eyert, D. Vollhardt, and V. I. Anisimov, Phys. Rev Lett., {\bf 86}, 5345 (2001),  G. Keller, K. Held, V. Eyert, D. Vollhardt, and V. I. Anisimov, Phys. Rev. B, \textbf{70}, 205116 (2004).

\bibitem{oka} A. Poteryaev, Phys. Rev. B, {\bf 76}, 085127 (2007).

\bibitem{notadop} Note that for the doped Hubbard model a much smaller variation of $SW$ is obtained (around 10\%) since in this case the metallic behavior is strengthened by the extra charge\cite{Toschi}.

\bibitem{merino} J. Merino and R. H. McKenzie, Phys. Rev. B {\bf 61}, 7996 (2000).

\bibitem{mcwhan69} D. B. McWhan,  T. M. Rice, and J. P. Remeika, Phys. Rev. Lett. \textbf{23}, 1384 (1969).

\bibitem{nota_prob} We emphasize that our calculation, though much more realistic than the idealized single-band Hubbard model, still involves some approximations. Namely we do not consider off-diagonal self-energies connecting different bands, and we do not compute the optical dipole matrix elements. Finally, the exchange interactions are approximated through an Ising-like term in order to solve the model through QMC. The latter approximation reduced the tendency to form local spin-1 states, so that the local spin entropy is the same as in the model by C. Castellani, C.R. Natoli and J. Ranninger (Phys. Rev. B {\bf 18}, 4945 (1978), ibid. {\bf 18}, 4967 (1978), ibid. {\bf 18}, 5001 (1978)).

\bibitem{panaccione} G. Panaccione, M. Altarelli, A. Fondacaro, A. Georges, S. Huotari, P. Lacovig, A. Lichtenstein, P. Metcalf, G. Monaco, F. Offi, L. Paolasini, A. Poteryaev, O. Tjernberg, and M. Sacchi,  Phys. Rev. Lett. {\bf 97}, 116401 (2006).

\bibitem{mlaad} M. Laad, L. Craco and E. MŸller-Hartmann, Phys. Rev. B. {\bf 73}, 045109 (2006).

\end{thebibliography}
\end{document}